# Betatron resonance electron acceleration and generation of quasi-monoenergetic electron beams using 200fs Ti:Sapphire laser pulses

D. Hazra[2,*], A. Moorti[1,2,†], B. S. Rao[1], A. Upadhyay[1], J. A. Chakera[1,2], and P. A. Naik[1,2]

[1]Laser Plasma Section, Raja Ramanna Centre for Advanced Technology, Indore-452013, India

[2]Homi Bhabha National Institute, Training School Complex, Anushakti Nagar, Mumbai 400094, India

*E-mail: dhazra@rrcat.gov.in
†E-mail: moorti@rrcat.gov.in

## Abstract

Generation of collimated, quasi-monoenergetic electron beams (peak energy ~17-22MeV, divergence ~10mrad, energy spread ~20%) by interaction of Ti:sapphire laser pulse of 200fs duration, focussed to an intensity of ~ $2.1 \times 10^{18}$ W/cm$^2$, with an under-dense (density~$3.6 \times 10^{19}$ to ~$1.1 \times 10^{20}$ cm$^{-3}$) He gas-jet plasma was observed. Two stages of self-focusing of the laser pulse in the plasma were observed. Two groups of accelerated electrons were also observed associated with these stages of the channeling and is attributed to the betatron resonance acceleration mechanism. This is supported by 2D PIC simulations performed using code EPOCH and a detailed theoretical analysis which shows that present experimental conditions are more favorable for betatron resonance acceleration and generation of collimated, quasi-thermal/quasi-monoenergetic electron beams.

Laser driven plasma based electron acceleration is seen as a potential candidate for development of future compact accelerators. In this regard, laser wake-field acceleration (LWFA) technique [1-5] has been successfully used for generation of high-quality quasi-monoenergetic (QM), near GeV and more than GeV energy electron beams using "bubble regime" [6-15] where a short duration laser pulse such that $L = c\tau \leq \lambda_p$ is used, where $L$ is laser pulse length, $c$ is speed of light, $\tau$ is FWHM (Full-width at half-maximum) pulse duration, and $\lambda_p$ is plasma wavelength. On the other hand, generation of QM electron beams of few tens of MeV energies have also been demonstrated by self-modulated laser wake-field acceleration (SM-LWFA) [16-23] at comparatively higher plasma densities ($L \gg \lambda_p$), by using laser pulses as long as 90fs, where it was suggested that small laser pulselets formed due to self-modulation could create bubble regime conditions [16]. The earlier reported experiments with comparatively longer laser pulses (few hundreds of fs to ps) have shown generation of relativistic electron beams through SM-LWFA but with broad continuous spectrum [24-27]. Another possible mechanism of betatron resonance acceleration (resonant transfer of energy from the laser field to the oscillating electrons in laser channel i.e. direct laser acceleration: DLA) in similar conditions was proposed by Pukhov *et al.* [28,29]. There is only one dedicated experimental report by Gahn *et al.* [30,31] from underdense gas jet plasma target and one theoretical report by Tsakiris *et al.* [32] on betatron resonance acceleration. In some other reports on DLA e.g. Mangles *et al.*, [33] and also in some reports on SM-LWFA [34-36] applicability of this mechanism has been discussed. Recently, generation of non-Maxwellian electron beams accelerated by DLA in near critical plasmas produced using few nm thick targets has also been reported [37].

In this letter, we report an experimental study on electron acceleration along with channeling of a ~200fs duration Ti:sapphire laser pulse, focussed to an intensity of

~$2.1\times10^{18}$ W/cm$^2$, in an under-dense He plasma produced from a 1.2mm long gas-jet target (plasma density~$3.6\times10^{19}$ to ~$1.1\times10^{20}$cm$^{-3}$). Although overall electron spectrum was quasi-thermal, generation of prominently two groups of electron beams was observed with peak energy in the range of 8-10MeV and 15-25MeV, with divergence ~10-20mrad. Further, in many shots, generation of single QM electron beams (peak energy~17-22MeV, energy spread~20%) was also observed. The two groups of accelerated electrons are associated with the observed two stages of laser self-focusing and channeling. 2D PIC simulations using code EPOCH [38] along with detailed theoretical analysis, and also PIC simulations reported by other groups in similar experimental conditions [28-30], suggest applicability of betatron resonance acceleration mechanism. Studies on betatron resonance electron acceleration in laser plasma interaction is desirable and of interest for betatron radiation generation associated with this particular regime of electron acceleration [39], and not much experimental work has been reported in this area. Further, characteristics of electron beams generated is also imporatnt, as it may be pointed out here that generation of QM electron beams using such long (200fs) laser pulses in underdense gas-jet plasma where dominant acceleration mechanism is betatron resonance acceleration have not been reported earlier.

The experiment was performed using a 150TW, 25fs Ti: sapphire laser system. The pulse duration was stretched to ~200fs (power ~15TW) by changing compressor gratings separation, thereby introducing a slight positive chirp in the laser pulse. The experimental set up used was similar to that used in our previous studies [23]. The laser pulse was focused using an f/5 off-axis parabola to a focal spot size of ~$25\times12$ $\mu$m$^2$, half width at 1/e$^2$ of the maximum, (peak intensity ~$2.1\times10^{18}$ W/cm$^2$, normalized laser vector potential $a_0$~1). The Rayleigh length ($Z_R$) was estimated to be ~180 $\mu$m. A supersonic He gas-jet target was used (slit nozzle: 1.2mm$\times$10mm) with plasma density of ~$3.6\times10^{19}$cm$^{-3}$ to

~$1.1\times10^{20}$cm$^{-3}$. DRZ phosphor screen was used as an electron detector and a magnetic spectrograph (diameter: 50mm, pole gap: 9mm, peak magnetic field: 0.45T) with a resolution of ~10% at 10MeV and ~20% at 20MeV (for a beam divergence of 10mrad) was used to measure electron energy. A 6mm thick aluminum plate placed before the phosphor screen blocked the laser and also electrons with energy below~2.7MeV. A $90^0$ side scattering imaging on a 14 bit CCD camera with 5X magnification was used to study the laser propagation.

Generation of relativistic electron beams was studied by varying the plasma density for a fixed laser intensity of ~$2.1\times10^{18}$ W/cm$^2$. Collimated electron beams in the forward direction were observed above a threshold plasma density of ~$3.6-4\times10^{19}$cm$^{-3}$ (Fig.1a). For density of up to ~$7\times10^{19}$cm$^{-3}$, similar electron beams with full angle divergence of ~40mrad (FWHM) were observed (Fig.1a (i) and (ii)). At higher density, the overall beam divergence increased (~120mrad), however an intense central spot was still seen (Fig.1a (iii)).

Electron spectrum recorded has a quasi-thermal distribution and showed generation of electrons with peak energy of ~8-10MeV. Most of the times, a second group of electrons with comparatively higher energy of ~15-25MeV was also seen. Typical electron beam spectra recorded for various plasma densities are shown in Fig.1b, and a typical raw image of spectrum for electron density of ~$1.1\times10^{20}$cm$^{-3}$ showing multiple groups of electron beam formation is shown in the inset of Fig.1b. Maximum electron energy extended upto ~30MeV (at 10% of the peak energy amplitude value). With increase in plasma density, slight increase in the peak energy of electrons was observed for both the group of electrons. Interestingly, in several shots, single highly-collimated (divergence ~5-10mrad) QM electron beam (peak energy ~17-22MeV, energy spread

~20%) was also observed (Fig.1c). Inset in Fig.1c shows a typical raw image of such spectra at a density of $8\times10^{19} cm^{-3}$.

Relativistic self-focusing and guiding of the laser pulse inside plasma was observed for the range of plasma density used. A typical laser created channel for plasma density of $6\times10^{19} cm^{-3}$ at a fixed laser power of 7.5TW is shown in Fig.1d. Initially as the laser pulse enters plasma, self-focusing of the laser is observed which slightly defocuses (bulging of the channel) for a small distance followed by a further self-focusing in the later stage of propagation. Average channel radius (FWHM) increases from initial ~4-5 $\mu$m to ~6-10 $\mu$m in the middle and then converges to ~4 $\mu$m at the end. The total laser-plasma interaction length observed was ~450 $\mu$m (~2.5$Z_R$). In some of the shots a slight bending in the later part of the laser channel was observed and associated with it two electron beams were observed on the phosphor screen (without magnet in the path) as shown in Fig.1e. Otherwise, for the straight channels single electron beam profiles as shown in Fig.1a were observed.

One of the possible mechanisms of electron acceleration in laser channels for $L > \lambda_p$ is SM-LWFA. Considering which for the highest density of $1.1\times10^{20} cm^{-3}$ dephasing limited maximum energy of the electrons is expected to be only ~5MeV. Also energy should decrease with density [5]. In contrary maximum energy upto ~30MeV, along with an increase in the electron energy with density was observed (Fig.1b). Further, earlier simulations carried out for the similar experimental conditions have shown that the wake-fields are effective only at the foot of the laser pulse [29,30], signifying dominant electron acceleration mechanism of betatron resonance acceleration [30]. Next, we have observed electron acceleration for high values of $P/P_c$ (9-28), consistent with earlier report suggesting that for betatron resonance acceleration $P/P_c$ should be greater than 6 [28]. The above discussion suggests that the SM-LWFA may not be applicable in the present case.

Moreover, observation of quasi-thermal (Fig.1b) and QM (Fig.1c) electron beams is also in contrast to the continuous electron spectrum obtained through SM-LWFA using hundreds of fs laser pulses [24-27]. During the present experimental campaign SM-LWFA regime was also achieved when the pulse duration was reduced to ~55fs showing stable generation of QM beams similar to our earlier observation [23], and various other reports [16-22].

To understand the applicable acceleration mechanism and to support experimental observations we performed a 2D PIC simulation using code EPOCH [38]. The simulation was modeled with a simulation box of 450μm× 60μm, grid size of λ/20 and 30 particles per cell. Laser pulse of 200fs duration and intensity of $2\times10^{18}$ W/cm$^2$, propagating along x-direction with polarization along y-direction enters the simulation box from left and interacts with plasma having a linear density ramp (0 to $n_e=7\times10^{19}$cm$^{-3}$) of 100μm followed by a constant density of $n_e$ for further 350μm. Fig.2a ((i)-(iii)) shows the electron density profile at different time steps in laser propagation direction which clearly shows channel formation associated with relativistic self-focusing. Immediately after entering, the plasma self-focusing of laser pulse occurs (for initial ~150μm plasma length with channel radius of ~4-5μm is observed (Fig.2a (i)). In the middle of the plasma channel defocusing of the laser pulse occurs (channel radius to ~8-10μm) as shown in Fig.2a (ii). Finally a second stage of laser self-focusing occurs (channel radius~4μm) during remaining part of ~150μm of laser propagation inside plasma (Fig. 2a (iii)). This is consistent with the experimentally recorded laser channels as shown in Fig.1d. For entire self-focused region of the laser propagation, generation of laser-driven wakefield is not observed, except at the middle of the propagation where slight defocusing and a modulation of the laser pulse is observed (Fig.2b), that too within a small portion at the front part of the laser pulse, (Fig.2c) and which also could not sustain in the later stage of propagation. Hence in the present case applicable mechanism could only be betatron resonance acceleration. Fig.2d ((i)-(iii)) shows $P_x$ (momentum) vs x (propagation distance) at different time steps corresponding to channels shown in Fig.2a. Corresponding to ~100-150µm of laser propagation in the two stages of self-focusing, we observe two stages of acceleration of electrons with ~18MeV in the first stage and enhancement of electron energy upto ~50MeV in the second acceleration stage along with formation of multiple electron bunches and appearance of quasi-monoenergetic features as marked by red circles in Fig.2d (ii) & (iii). This is consistent with the observed two groups of electrons with quasi-monoenergetic feature experimentally. Two groups of electrons are associated with the above described two stages of the laser channeling. The first group of electrons with peak energy of ~8-10MeV (Fig.1b) is accelerated in the first stage of the laser channel. Higher energy electrons (~15-25MeV) are generated due to acceleration in the later part. Mangles *et al.*, [33] have also reported through simulation two stages of electron acceleration in case of DLA where enhancement in the electron energy was attributed to the magnetically constricted part of the laser channel in the later stage of the propagation. Observation of two electron beams associated with bending of laser channel in the later part of laser propagation as shown in Fig.1e above also supports this process.

Further, we plot the variation of normalized maximum transverse laser field ($eE_y/mc\omega$) and longitudinal wakefield ($eE_x/mc\omega_p$) along the propagation distance (Fig.3a). This shows that during the entire propagation length laser field is higher compared to the wakefield. Separate contribution of the maximum transverse ($\gamma_y$) and longitudinal ($\gamma_x$) energy gain ($\gamma_y = -\int_0^t \frac{2ep_y E_y}{(mc)^2} dt, \gamma_x = -\int_0^t \frac{2ep_x E_x}{(mc)^2} dt$) to the total energy gain $\gamma^2 = 1 + \gamma_x + \gamma_y$ were studied and plotted along the propagation distance (Fig.3b).

This shows that $\gamma_y$ from the laser field (DLA) is much greater than $\gamma_x$ and also the gain is significantly higher in the later stage of propagation. The variation of $\gamma_y$ with total $\gamma$ follows a straight line with slope close to 1, thereby emphasizing the dominant contribution of DLA over wakefield to the total energy gain of electrons inside the channels (Fig.3c). Increase in the longitudinal momentum with distance also shows electrons oscillating in the laser field gain transverse momentum by resonant transfer of energy from the field, which is converted to longitudinal direction via v ×B force. Our simulation results are consistent with the previous simulations reported by Gahn *et al.* [30] in similar experimental conditions and also Mangles *et al.* [33], both suggesting betatron resonance acceleration mechanism of electrons.

Next, we performed a detailed theoretical analysis of betatron resonance acceleration in our experimental conditions to support our observations. In betatron resonance acceleration energy of the accelerated electrons $\gamma$ with phase $\phi$ is given by [32]:

$$\frac{d\gamma}{d\phi} = -\frac{eA_0 v_{xA} \cos\phi}{2mc^2 \left(\omega - \frac{\omega_{b0}}{\sqrt{\gamma}} - kv_z\right)} \tag{1}$$

Integrating Eq. (1) we get $F(\gamma) = -P\sin\phi + C_1$, where F($\gamma$) is given by,

$$F(\gamma) = \gamma - \eta\sqrt{(1-\alpha_0)\gamma^2 - 1} + \eta\cos^{-1}\frac{1}{\gamma\sqrt{1-\alpha_0}} - 2\gamma^{1/2}\frac{\omega_{b0}}{\omega} \tag{2}$$

Here $A_0$ is the electric field amplitude, $v_{xA}$ and m is the on axis velocity and rest mass of the electrons, $\omega$ is the laser frequency, $\omega_{b0} = \omega_b \sqrt{\gamma}$ represents the bounce frequency of the oscillation, $k = (\omega/c)\eta$ is the wave number, $\eta = \left(1 - \omega_p^2/\omega^2 \left(1 + a_0^2/2\right)^{-1/2}\right)^{1/2}$ is the ratio of group velocity of the laser in plasma to that in vacuum, $v_z = c(1 - 1/\gamma^2 - v_{xA}^2/2c^2)^{1/2}$ is the axial velocity of electron, $P = a_0 v_{xA}/2c$, $C_1 = F(\gamma_0) + P\sin\phi_0$ is the integration constant, $\gamma_0$ is the initial energy of electrons, and $\phi_0$ is the initial phase of the wave seen by electron, and $\alpha_0 = v_{xA}^2/2c^2$. Fig.4a shows the plot of $F(\gamma)$ vs $\gamma$ for $\eta = 0.983$ (average value for the density range used).

Next, separatrix ($\gamma$ vs $\phi$) was plotted (Fig.3b) using equation [32]:

$$F(\gamma) - F_{min} = P(1 - \sin\phi) \quad (3)$$

Where $F_{min}$ is the minimum value of F obtained from Fig.4a. The largest value of the right hand side of the above equation is equal to 2P. Therefore, a horizontal line was drawn in Fig.4a at a height of 2P from $F_{min}$, which cuts the $F(\gamma)$ curve at two points R and S which corresponds to the bottom and top of the separatrix (Fig.4b) respectively. The cross points of the separatrix occur at $\gamma = \gamma_{opt}$. Fig.4a and 4b show that the maximum energy acquired by a trapped electron is~34MeV ($\gamma(S) \sim 67.5$), close to the energy observed in the present experiment.

Further, the transverse energy gain $\gamma_T$ of the electrons in the combined fields of the laser and the static fields of the laser channel, is given by [32]:

$$\gamma_T = 1/(1-\eta^2)[-Q\eta + (1 + Q^2 - \eta^2)^{(1/2)}] \quad (4)$$

$$\text{where,} \quad Q = \eta\left[\left(1 + \frac{a_0^2}{2}e^{-x_T^2/r_0^2}\right)^{\frac{1}{2}} - \left(1 + \frac{a_0^2}{2}\right)^{\frac{1}{2}} + \frac{\omega_c \pi r_0}{\omega \eta \lambda}\left(e^{-x_T^2/r_0^2} - 1\right) - \frac{1}{\eta}\sqrt{\gamma_0^2 - 1} + \gamma_0\right] \quad (5)$$

Here, $x_T$ is the turning point radius (the maximum transverse amplitude of oscillation), $r_0$ is laser focal spot, $\omega_c$ is the cyclotron frequency, and $\lambda$ is laser wavelength. In Fig.3c we have plotted $\gamma_T$ vs $x_T/r_0$ showing significant energy gain from lower electron oscillation amplitude in our case compared to the other previous reports [30,32,33]. This suggests that in the present experimental conditions significant energy gain occurs from regions very close to the laser channel axis, where the laser intensity also remains constant to its peak value. This is because of comparatively larger laser focal spot used in the present experiment. Hence, our experimental conditions are more favorable for generating collimated electron beams through betatron resonance acceleration mechanism.

Finally, we discuss the observation of quasi-thermal/QM features by estimating the dephasing length ($L_d$) using equations:

$$\frac{d\gamma}{d\xi} = -\frac{a_0 \left(\alpha_0/2\right)^{\frac{1}{2}}}{\left(1-\alpha_0 - 1/\gamma^2\right)^{\frac{1}{2}}} \cos\phi \qquad (6)$$

$$\frac{d\phi}{d\xi} = \frac{1 - \omega_{b0}/\omega\gamma^{\frac{1}{2}}}{\left(1-\alpha_0 - 1/\gamma^2\right)^{\frac{1}{2}}} - \eta \qquad (7)$$

Here, $\xi = z\omega/c$ and z is the propagation distance. $L_d$ was estimated by plotting $\gamma$ vs z with initial conditions at $\xi=0$, $\phi = \pi/2 + \Delta$, where $\Delta = 0.02$, and $\gamma_0 = \gamma_{opt}$. For the range of plasma density used $L_d$ was in the range of 100-200 $\mu$m (as shown in Fig.4d for $\eta = 0.983$). The maximum channel length of ~450 $\mu$m was observed in the present experiment, but as discussed above, the acceleration occurred in the two stages of laser channeling and observed channel lengths of ~100-150μm in both the stages are comparable to $L_d$, hence leading to bunching of the trapped electrons and appearance of quasi-thermal/QM features in the electron beam spectra. Gahn *et al.* [30] and recently Toncian *et al.* [37] have reported observation of quasi-thermal/non-Maxwellian energy distribution through DLA.

In conclusion, we have observed collimated (≤40mrad), relativistic electron beams of up to ~30MeV energy by betatron resonance acceleration mechanism through propagation of a 200fs duration Ti:Sapphire laser pulse in under-dense (plasma density~$3.6\times10^{19}$ to ~$1.1\times10^{20}$cm$^{-3}$) helium plasma target. Generation of QM ($\Delta E/E$~10-20%) electron beams with peak energy of ~17-22MeV was also observed with such long laser pulses. 2D PIC simulations showed that dominant acceleration mechanism of electrons is DLA. Further, our observations were supported and explained using detailed theoretical analysis of betatron resonance acceleration mechanism. The study could be useful towards betatron radiation generation [40,41] in laser plasma channel associated with direct laser accelerated electrons [38].

**Acknowledgements:**

The authors would like to acknowledge the support provided by R. A. Khan, and A. Singla for the laser operation, D. Karmakar for help in setting up the experiment, and R. P. Kushwaha, S. Sebastin, and K. C. Parmar for providing mechanical / workshop support.

**References**

[1] T. Tajima and J. M. Dawson, Phys. Rev. Lett. **43**, 267 (1979).

[2] S. P. D. Mangles, C. D. Murphy, Z. Najmudin, A. G. R. Thomas, J. L. Collier, A. E. Dangor, E. J. Divall, P. S. Foster, J. G. Gallacher, C. J. Hooker, D. A. Jaroszynski, A. J. Langley, W. B. Mori, P.A. Norreys, F. S. Tsung, R. Viskup, B. R. Walton, and K. Krushelnick, Nature, **431**, 535 (2004).

[3] C. G. R. Geddes, Cs. Toth, J. van. Tilborg, E. Esarey, C. B. Schroeder, D. Bruhwiler, C. Nieter, J. Cary, and W. P. Leemans, Nature, **431**, 538 (2004).

[4] J. Faure, Y. Glinec, A. Pukhov, S. Kiselev, S. Gordienko, E. Lefebvre, J.-P. Rousseau, F. Burgy, and V. Malka, Nature, **431**, 541 (2004).

[5] E. Esarey, C. B. Schroeder, and W. P. Leemans, Rev. Mod. Phys. **81**, 1229 (2009).

[6] A. Pukhov and J.Meyer-ter-Vehn, Appl. Phys. B: Lasers Opt.**74**, 355 (2002).

[7] W. Lu, M. Tzoufras, C. Joshi, F. S. Tsung, W. B. Mori, J. Vieira, R. A. Fonseca, and L. O. Silva, Phys. Rev. ST Accel. Beams **10**, 061301 (2007).

[8] W. P. Leemans, A. J. Gonsalves, H.-S. Mao, K. Nakamura, C. Benedetti, C. B. Schroeder, Cs. Tóth, J. Daniels, D. E. Mittelberger, S. S. Bulanov, J.-L. Vay, C. G. R. Geddes and E. Esaray, Phys. Rev. Lett. **113**, 245002 (2014).

[9] X. Wang, R. Zgadzaj, N. Fazel, Z. Li, S. A. Yi, X. Zhang, W. Henderson, Y.-Y. Chang, R. Korzekwa, H.-E. Tsai, C.-H. Pai, H. Quevedo, G. Dyer, E. Gaul, M. Martinez, A. C. Bernstein, T. Borger, M. Spinks, M. Donovan, V. Khudik, G. Shvets, T. Ditmire, and M. C. Downer, Nat. Commun. **4**, 1988 (2013).

[10] H. T. Kim, K. H. Pae, H. J. Cha, I. J. Kim, T. J. Yu, J. H. Sung, S. K. Lee, T. M. Jeong, and J. Lee, Phys. Rev. Lett. **111**, 165002 (2013).

[11] W. P. Leemans, B. Nagler, A. J. Gonsalves, Cs. Tóth, K. Nakamura, C. G. R. Geddes, E. Esaray, C. B. Schroeder, and S. M. Hooker, Nat. Phys. **2**, 696 (2006) .

[12] D. H. Froula, C. E. Clayton, T. Döppner, K. A. Marsh, C. P. J. Barty, L. Divol, R. A. Fonseca, S. H. Glenzer, C. Joshi, W. Lu, S. F. Martins, P. Michel, W. B. Mori, J. P.

Palastro, B. B. Pollock, A. Pak, J. E. Ralph, J. S. Ross, C. W. Siders, L. O. Silva, and T. Wang, Phys. Rev. Lett. **103**, 215006 (2009).

[13] S. Kneip, S. R. Nagel, S. F. Martins, S. P. D. Mangles, C. Bellei, O. Chekhlov, R. J. Clarke, N. Delerue, E. J. Divall, G. Doucas, K. Ertel, F. Fiuza, R. Fonseca, P. Foster, S. J. Hawkes, C. J. Hooker, K. Krushelnick, W. B. Mori, C. A. J. Palmer, K. TaPhuoc, P. P. Rajeev, J. Schreiber, M. J. V. Streeter, D. Urner, J. Vieira, L.O. Silva, and Z. Najmudin, Phys. Rev. Lett. **103**, 035002 (2009).

[14] S. Banerjee, N. D. Powers, V. Ramanathan, I. Ghebregziabher, K. J. Brown, C. M. Maharjan, S. Chen, A. Beck, E. Lefebvre, S. Y. Kalmykov, B. A. Shadwick, and D. P. Umstadter, Phys. Plasmas **19**, 056703 (2012).

[15] C. E. Clayton, J. E. Ralph, F. Albert, R. A. Fonseca, S. H. Glenzer, C. Joshi, W. Lu, K. A. Marsh, S. F. Martins, W. B Mori, A. Pak, F. S. Tsung, B. B. Pollock, J. S. Ross, L. O. Silva, and D. H.Froula, Phys. Rev. Lett. **105**, 105003 (2010).

[16] B. Hidding, K.-U. Amthor, B. Liesfeld, H. Schwoerer, S. Karsch, M. Geissler, L. Veisz, K. Schmid, J. G. Gallacher, S. P. Jamison, D. Jaroszynski, G. Pretzler, and R. Sauerbrey, Phys. Rev. Lett. **96**, 105004 (2006).

[17] E. Miura, K. Koyama, S. Kato, N. Saito, M. Adachi, Y. Kawada, T. Nakamura, and M. Tanimoto, Appl. Phys. Lett. **86**, 251501 (2005).

[18] A. Yamazaki, H. Kotaki, I. Daito, M. Kando, S. V. Bulanov, T. Zh. Esirkepov, S. Kondo, S. Kanazawa, T. Homma, K. Nakajima, Y. Oishi, T. Nayuki, T. Fujii, and K. Nemoto, Phys. Plasmas **12**, 093101 (2005).

[19] T. Hosokai, K. Kinoshita, T. Ohkubo, A. Maekawa, M. Uesaka, A. Zhidkov, A, Yamazaki, H. Kotaki, M. Kando, K. Nakajima, S. V. Bulanov, P. Tomassini, A. Giulietti, and D. Giulietti, Phys. Rev. E **73**, 036407 (2006).

[20] C.-T. Hsieh, C.-M. Huang, C.-L. Chang, Y.-C. Ho, Y.-S. Chen, J.-Y. Lin, J. Wang, and S.-Y. Chen, Phys. Rev. Lett. **96**, 095001 (2006).

[21] M. Mori, M. Kando, I. Daito, H. Kotaki, Y. Hayashi, A. Yamazaki, K. Ogura, A. Sagisaka, J. Koga, K. Nakajima, H. Daido, S. V. Bulanov, and T. Kimura , Phys. Lett. A **356**, 146 (2006).

[22] D. Z. Li, W. C. Yan, L. M. Chen, K. Huang, Y. Ma, J. R. Zhao, L. Zhang, N. Hafz, W. M. Wang, J. L. Ma, Y. T. Li, Z. Y. Wei, J. Gao, Z. M. Sheng, and J. Zhang, Opt. Express**22**, 12836 (2014).

[23] B. S. Rao, A. Moorti, R. Rathore, J. A. Chakera, P. A. Naik, and P. D. Gupta, Phys. Rev. ST Accel. Beams **17**, 011301 (2014).

[24] A. Modena, Z. Najmudin, A. E. Dangor, C. E. Clayton, K. A. Marsh, C. Joshi, V. Malka, C. B. Darrow, C. Danson, D. Neely, and F. N. Walsh, Nature (London) **377**, 606 (1995).

[25] D. Umstadter, S.-Y. Chen, A. Maksimchuk, G. Mourou, and R. Wagner, Science **273**, 472 (1996).

[26] R. Wagner, S.-Y. Chen, A. Maksimchuk, and D. Umstadter, Phys. Rev. Lett. **78**, 3125 (1997).

[27] D. Gordon, K. C. Tzeng, C. E. Clayton, A. E. Dangor, V. Malka, K. A. Marsh, A. Modena, W. B. Mori, P. Muggli, Z. Najmudin, D. Neely, C. Danson, and C. Joshi, Phys. Rev. Lett. **80**, 2133 (1998).

[28] A. Pukhov, Z.-M. Sheng, and J. Meyer-ter-Vehn, Phys. Plasmas **6**, 2847 (1999).

[29] A.Pukhov  and J. Meyer-ter-Vehn, Phys.Plasmas**5**, 1880 (1998).

[30] C. Gahn, G. D. Tsakiris, A. Pukhov, J. Meyer-ter-Vehn, G. Pretzler, P. Thirolf, D. Habs, and K. J. Witte, Phys. Rev. Lett. **83**, 4772 (1999).

[31] C. Gahn, G. D. Tsakiris, G. Pretzler, K. J. Witte, P. Thirolf, D. Habs, C. Delfin and C.-G. Wahlström, Phys. Plasmas **9**, 987 (2002).

[32] G. D. Tsakiris, C. Gahn, and V. K. Tripathi, Phys. Plasmas **7**, 3017 (2000).

[33] S.P.D. Mangles, B. R. Walton, M. Tzoufras, Z. Najmudin, R. J. Clarke, A. E. Dangor, R. G. Evans, S. Fritzler, A. Gopal, C. Hernandez-Gomez, W. B. Mori, W. Rozmus, M. Tatarakis, A. G. R. Thomas, F. S. Tsung, M. S. Wei, and K. Krushelnick, Phys. Rev. Lett. **94**, 245001 (2005).

[34] S. Masuda and E. Miura, Phys. Plasmas **16**, 093105 (2009).

[35] M. Adachi, E. Miura, S. Kato, K. Koyama, S. Masuda, T. Watanabe, H. Okamoto, A. Ogata, and M. Tanimoto, Japanese Journal of Applied Physics **45**, 4214 (2006).

[36] V. Malka, J. Faure, J. R. Marquès, F. Amiranoff, J. P. Rousseau, S. Ranc, J. P. Chambaret, Z. Najmudin, B. Walton, P. Mora, and A. Solodov, Phys. Plasmas **8**, 2605 (2001).

[37] T. Toncian, C. Wang, E. McCary, A. Meadows, A. V. Arefiev, J. Blakeney, K. Serratto, D. Kuk, C. Chester, R. Roycroft, L. Gao, H. Fu, X. Q. Yan, J. Schreiber, I. Pomerantz, A. Bernstein, H. Quevedo, G. Dyer, T. Ditmire and B. M. Hegelich, Matter and Radiation at Extremes **1**, 82 (2016).

[38] T. D. Arber, and K. Bennett, and C. S. Brady, C S and A. Douglas Lawrence, M. G. Ramsay, N. J. Sircombe, P. Gillies, R. G. Evans, H. Schmitz, A. R. Bell, and C. P. Ridgers, Plasma Physics and Controlled Fusion, **57**, 11, 1-26 (2015).

[39] T. W. Huang, A. P. L. Robinson, C. T. Zhou, B. Qiao, B. Liu, S. C. Ruan, X. T. He, and P. A. Norreys, Phys. Rev. E 93, 063203 (2016).

[40] S. Corde, K. Ta Phuoc, G. Lambert, R. Fitour, V. Malka, and A. Rousse, Rev. Mod. Phys. **85**, 1 (2013).

[41] S. Cipiccia, M. R. Islam, B. Ersfeld, R. P. Shanks, E. Brunetti, G. Vieux, X. Yang, R. C. Issac, S. M. Wiggins, G. H. Welsh, M.-P. Anania, D. Maneuski, R. Montgomery, G. Smith, M. Hoek, D. J. Hamilton, N. R. C. Lemos, D. Symes, P. P. Rajeev, V. O. Shea, J. M. Dias, and D. A. Jaroszynski, Nat. Phys. **7**, 867 (2011).

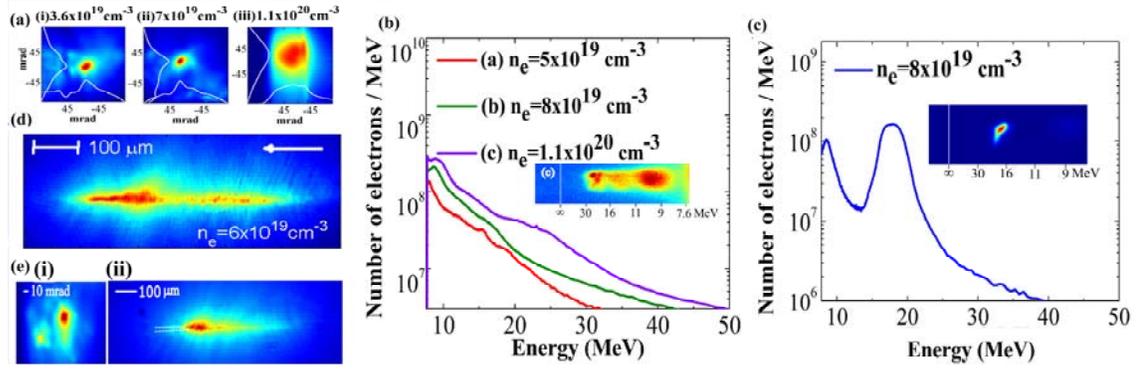

FIG.1. Experimental results. (a) Typical electron beam profiles (white curves shows the lineouts in horizontal and vertical directions). (b) Quasi-thermal electron beam spectra at different densities. Inset shows the raw image of a typical spectrum at $1.1\times10^{20}$ cm$^{-3}$ showing multiple groups of electron beam formation. (c) Quasi-monoenergetic electron beam spectrum. Inset shows a typical raw image. (d) A typical laser channel recorded at $6\times10^{19}$ cm$^{-3}$. White arrow shows laser propagation direction. (e) Observation of two electron beams (Frame-i) due to bending of laser channel in the later part of propagation (Frame-ii). Density: $5\times10^{19}$ cm$^{-3}$. White dotted lines show axes of two stages of laser propagation.

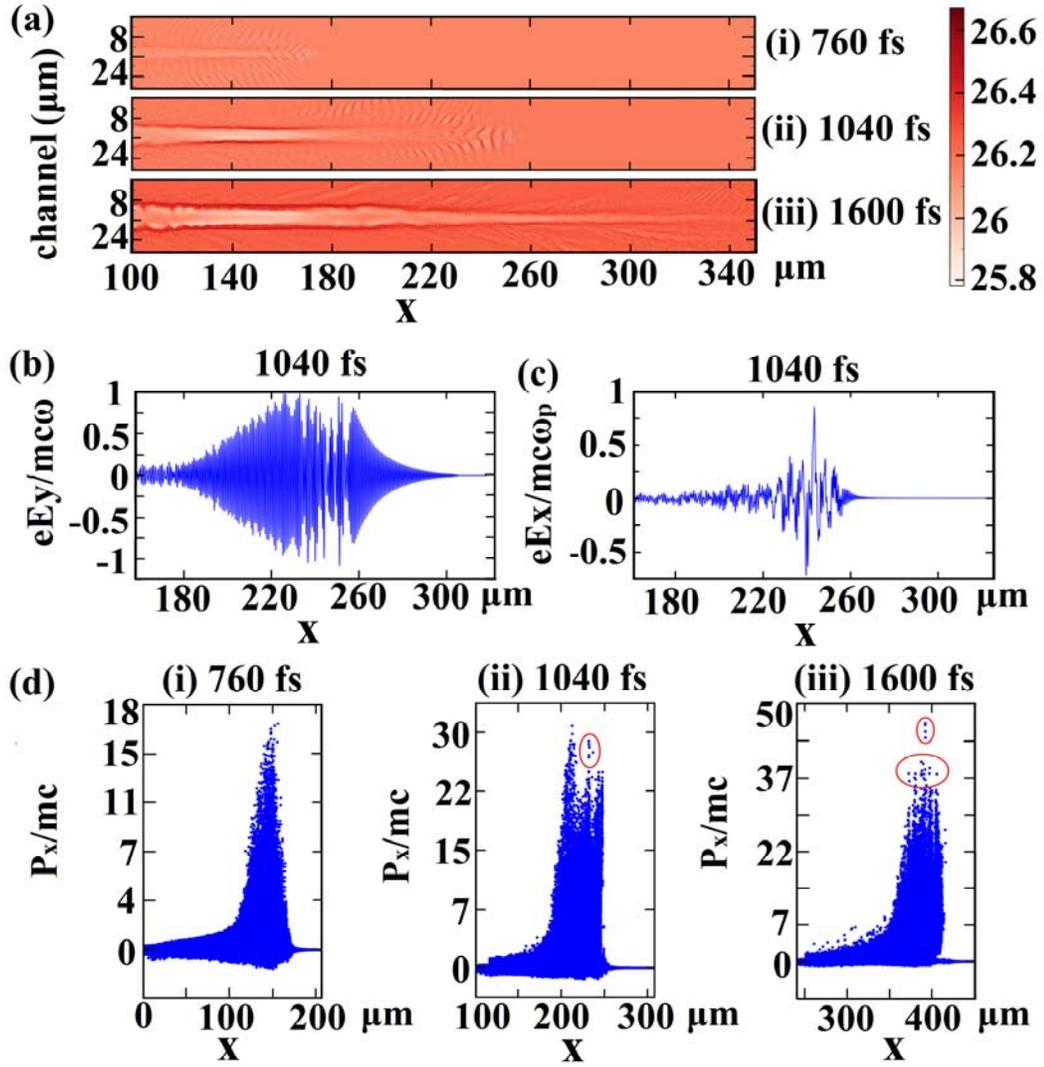

FIG.2. 2D PIC simulation results using EPOCH code. (a) Electron density profiles along x at different time steps (i) 760 fs, (ii) 1040 fs, (iii) 1600 fs. (b) Normalized laser electric field along x at 1040 fs. (c) Normalized longitudinal wakefield along x at 1040 fs. (d) Normalized $P_x$ vs x at different time steps (i) 760 fs, (ii) 1040 fs, (iii) 1600 fs. Red circles in (ii) & (iii) mark the appearance of quasi-monoenergetic features.

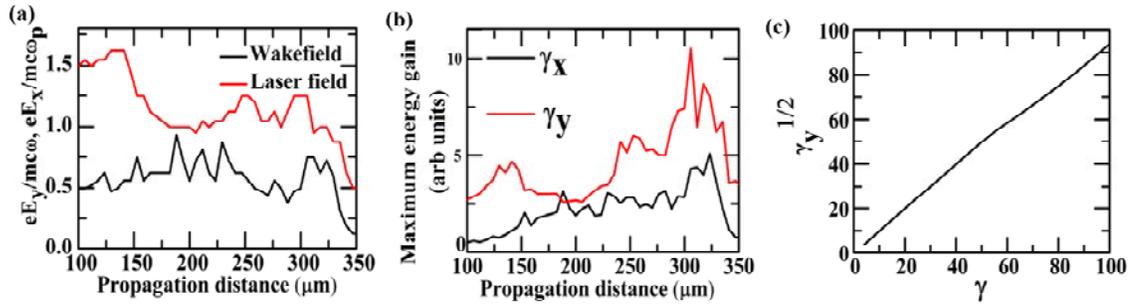

FIG.3. (a) Variation of laser electric field and longitudinal wakefield with propagation distance. (b) Variation of maximum transverse ($\gamma_y$) and longitudinal ($\gamma_x$) energy gain along propagation direction. (c) Plot of $\gamma_y$ vs $\gamma$.

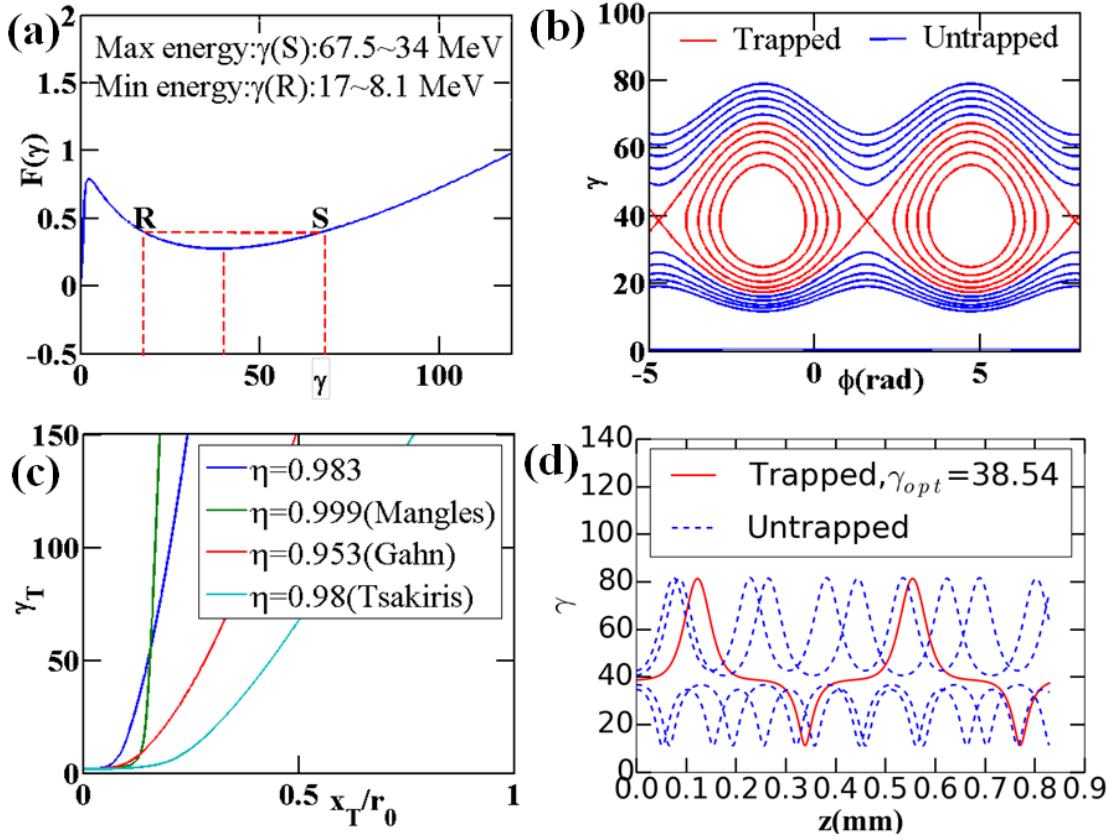

FIG.4. Analytical results. (a) F($\gamma$) vs $\gamma$ plot. (b) Separatrix of trapped (red) and untrapped (blue) electrons. (c) Transverse energy gain ($\gamma_T$) vs $x_T/r_0$ for conditions of Tsakiris [32] ($a_0=3$, $\eta=0.98$, $\gamma_0=1.81$ and $r_0=1.6\mu m$), Gahn [30] ($a_0\sim1.42$, $\eta=0.953$, $\gamma_0=1.81$ and $r_0=7.5\mu m$), Mangles [33] ($a_0=15$, $\eta=0.999$, $\gamma_0=1.81$ and $r_0=3\mu m$) and present experimental conditions ($a_0\sim1$, $\eta=0.983$, $\gamma_0=1.81$ and $r_0=12.5\mu m$). (d) Energy gain of electrons ($\gamma$) vs propagation distance (z) of trapped (red solid line) and untrapped (blue dashed lines) electrons. Other parameters for present case: $\omega_{b0}/\omega=0.2$, $\alpha_0=0.03$ and $P=0.0625$.